\documentstyle[twocolumn,aps]{revtex}

\input psfig.tex

\newcommand{\ltrsim}{\mathrel{\lower .3ex \rlap{$\sim$}
\raise .5ex\hbox{$<$}}}
\newcommand{\gttrsim}{\mathrel{\lower .3ex \rlap{$\sim$}
\raise .5ex\hbox{$>$}}}

\begin{document}
\draft

\twocolumn[\hsize\textwidth\columnwidth\hsize\csname %
@twocolumnfalse\endcsname

\title{
Does the Two-Dimensional $t$-$J$ Model have Hole Pockets?
}

\author{
W. O. Putikka$^{a,b*}$, M. U. Luchini$^c$ and R. R. P. Singh$^d$
}

\address{
$^a$Department of Physics, University of Cincinnati, Cincinnati, OH 45221-0011\\
$^b$Department of Physics, The Ohio State University, Mansfield, OH 44906\\
$^c$Department of Mathematics, Imperial College, London SW7 2BZ, United
Kingdom\\
$^d$Department of Physics, University of California, Davis, CA 95616\bigskip\\
}

\maketitle
\begin{abstract}
We have calculated the high temperature series for the momentum 
distribution function $n_{\bf k}$ of the 2D $t$-$J$ model to 
twelfth order in inverse temperature.  
By extrapolating the series to $T=0.2 J$ we investigate the
possibility of hole pockets in the $t$-$J$ model.  We find no
indication of hole pockets at an electron density of $n=0.9$ 
with $J/t=0.5$ or $J/t=1.0$.\\
\vspace{0.3in}
\end{abstract}
]

The locus in momentum space of low energy single particle
excitations is of prime importance for what it can tell us
about the nature
of the low energy degrees of freedom of the two dimensional $t$-$J$
model.  Two alternatives have been widely discussed: a) a large
Fermi surface consistent with Luttinger's theorem for a nearly half
filled band and b) small hole pockets centered around
$(\pi/2, \pi/2)$, similar to a lightly doped valence band in a
semiconductor\cite{palee}.  From the available experimental data 
which of these 
is correct for high temperature superconductors, if either, is still 
an open question.  Angle resolved photoemission experiments and
neutron scattering experiments generally support a large Fermi
surface.  However, transport measurements are more consistent with
a small number of positive charge carriers, in line with b) above.
The motivation for hole pockets centered around $(\pi/2, \pi/2)$
came from theoretical studies of a single hole in a N\'eel 
background\cite{palee}.  
More recently, hole pockets have been discussed 
theoretically\cite{kampf,duffy} and
experimentally\cite{aebi} in regard to shadow bands.  

A detailed investigation of hole pockets requires the 
complete single particle spectral function.  However,
Duffy and Moreo\cite{duffy} have discussed the signature of hole
pockets in $n_{\bf k}$ for a spin density wave mean-field approximation
to the Hubbard model\cite{kampf}.  Within this approach a deep notch
centered on $(\pi/2,\pi/2)$ develops in $n_{\bf k}$ for low enough
temperatures.  Below we investigate the possibility
of this type of hole pocket in the 2D $t$-$J$ model.

To investigate the nature of the occupied states for the 2D $t$-$J$
model we calculated the high temperature series for the momentum
distribution $n_{\bf k}$ to twelfth order in inverse temperature.
This is an extension of an earlier series calculation by Singh
and Glenister\cite{singh} to eighth order in inverse temperature.  
Our series
coefficients agree with theirs through eighth order.
The Hamiltonian for the $t$-$J$ model is given by
\begin{equation}
H=-t\sum_{\langle ij\rangle,\sigma}\left(c_{i\sigma}^{\dagger}
c_{j\sigma} + c_{j\sigma}^{\dagger}c_{i\sigma}\right)+J\sum_{
\langle ij\rangle}{\bf S}_i\cdot{\bf S}_j,
\end{equation}
with the constraint of no double occupancy.  The definition of the
single spin momentum distribution function is
\begin{equation}
n_{\bf k}=\sum_{\bf r}n_{\bf r}{\rm e}^{i{\bf k}\cdot{\bf r}},
\end{equation}
with $n_{\bf r}=\langle c_{0\sigma}^{\dagger}c_{{\bf r}\sigma}\rangle$.
The series coefficients are generated in coordinate space with the
Fourier transform to momentum space done exactly.  The resulting
series coefficients in momentum space are exact functions of the
wave vector.  By extrapolating the series to low temperatures and
using the exact momentum dependence of the series we 
can probe the momentum dependence of $n_{\bf k}$ with high precision.

To reach low temperatures we need to analytically continue the series
for $n_{\bf k}$.  A standard way to do this is to use Pad\'e
approximants.  For $n_{\bf k}$ the straightforward application of
Pad\'es does not work very well.  One way to improve the convergence of
Pad\'e approximants is to change the expansion variable to move the
poles of the Pad\'es  farther from the origin\cite{singh}.  
Exactly what change of
variable to choose is difficult to know for unknown functions.  However,
if the Pad\'es converge well for a certain choice of parameters
for $n_{\bf k}$, we can
calculate the ratio of $n_{\bf k}$ for this choice of parameters to 
$n_{\bf k}$ for
another choice of parameters, generally close by.  This gives a much 
simpler function to extrapolate with the Pad\'es and much better convergence.  
Once the extrapolation is done the desired result can
be obtained by multiplying the ratio by the known function.
For $n_{\bf k}$ it's easiest to fix $n$, $J$ and $t$ and vary {\bf k}.
This is the method of extrapolation used to obtain the data below.

Hole pockets are expected to be most prominent for the underdoped
cuprates\cite{palee}, $\delta \ltrsim 15\%$.  
To check this we fix the electron
density as $n=0.9$ ($\delta=10\%$).  The coupling constant ratio is
less well known, but is typically chosen to be around $J/t\sim0.4$.
As a representative value we use $J/t=0.5$.  Since  strong
antiferromagnetic correlations are expected to promote hole pocket
formation we have also investigated $n_{\bf k}$ for $J/t=1.0$.
The results of our calculations are shown in Fig. 1.  The momentum
distributions for the two values of $J/t$ are quite similar, with
$n_{\bf k}$ slightly larger near $(\pi,\pi)$ for $J/t=1.0$.  Since
for both values of $J/t$ $n_{\bf k}$ must satisfy the sum rule
$\sum_{\bf k}n_{\bf k}=n/2$ there must be a corresponding redistribution
of weight for $n_{\bf k}$ at $J/t=1.0$ elsewhere in the Brillouin zone.  To check
this would require a more complete set of data than presented here.
The main feature of both curves presented in Fig. 1 is that $n_{\bf k}$
for the parameters chosen here is a smooth function of {\bf k}.
Our results also satisfy the inequality\cite{stephan}
$n_{\bf k}\le(1+\delta)/2$, with
$n_{\bf k}=0.55$ for ${\bf k}=0$ and $n_{\bf k}$ monotonically
falling with increasing {\bf k}\cite{peter}.

From the smooth, monotonic {\bf k} dependence of $n_{\bf k}$ there is
clearly no indication of hole pockets near $(\pi/2,\pi/2)$.  There is 
also no obvious sign at $T=0.2J$ of a quasiparticle discontinuity. 
Where to look for a quasiparticle discontinuity is a subtle problem
when $T\ne 0$\cite{randeria}.  
From Fig. 1 we can see that $n_{\bf k}=1/2$ is at a 
different wave vector than where $n_{\bf k}$ has its maximum gradient,
so we do not obtain a unique {\bf k}$_{\scriptscriptstyle F}$.  This 
problem will be addressed in a future publication\cite{putikka}.
\begin{figure}[ht]
\centerline{\psfig{figure=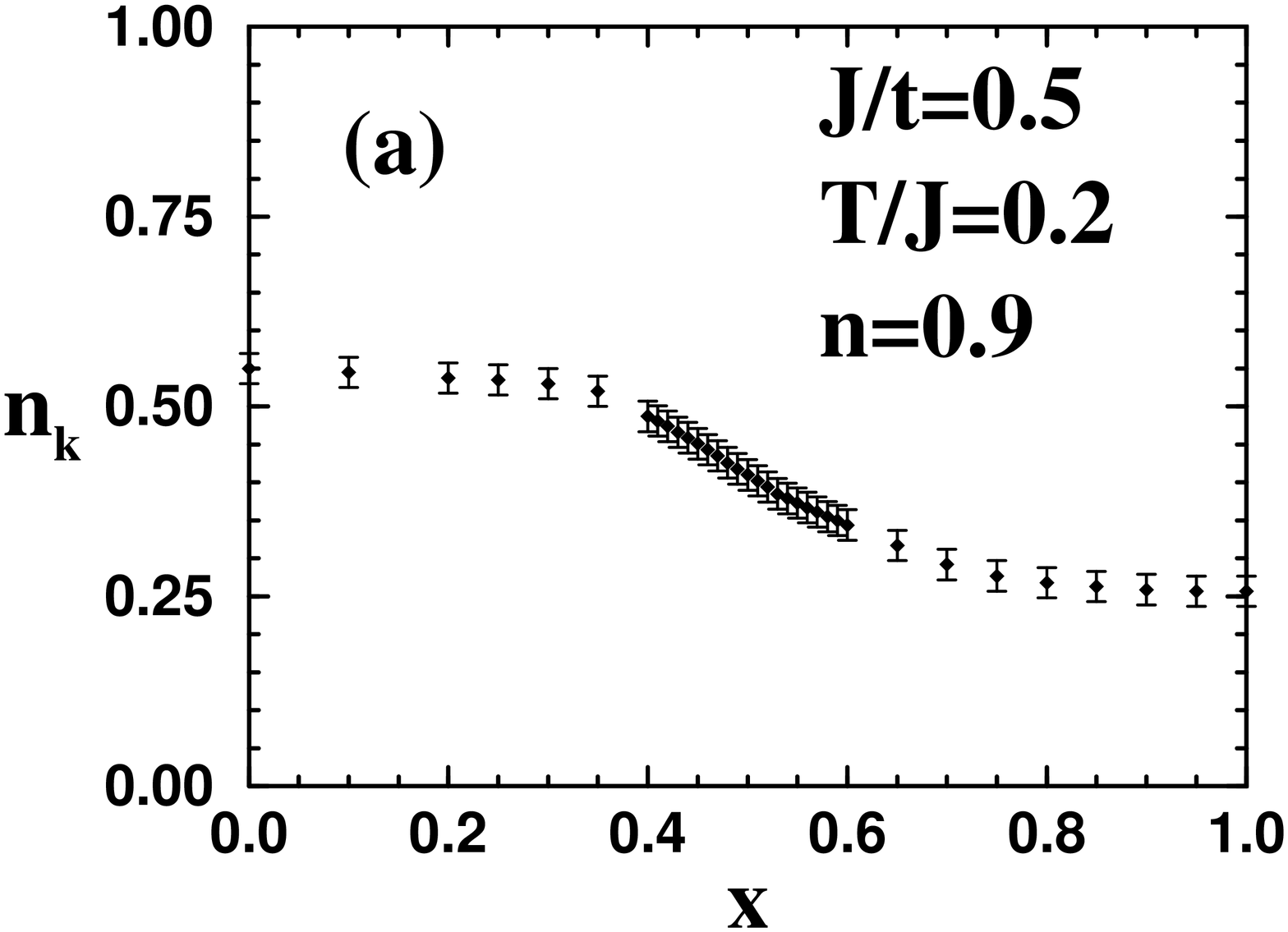,height=2in,angle=-90}}
\vspace{0.1in}
\centerline{\psfig{figure=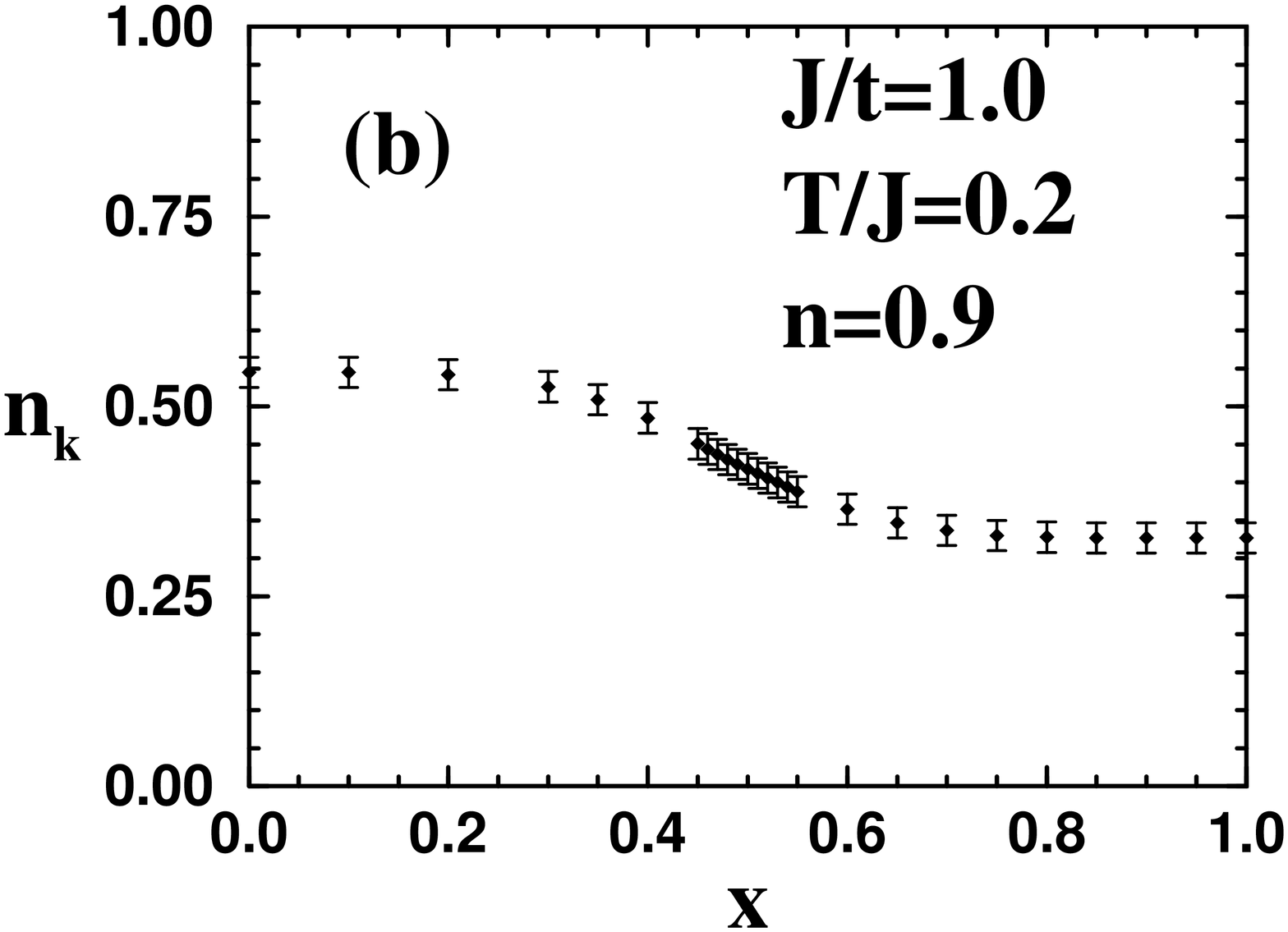,height=2in,angle=-90}}
\vspace{0.1in}
\caption{The momentum distribution $n_{\bf k}$ plotted versus wave vector
${\bf k}=({\rm x}\pi,{\rm x}\pi)$ along the diagonal of the square
Brillouin zone.  The electron density is fixed at $n=0.9$ and the
extrapolated temperature is $T/J=0.2$ (room temperature).
a) $J/t=0.5$ b) $J/t=1.0$.}
\end{figure}

The momentum distribution for strongly correlated 2D electrons has
been studied previously by a range of numerical methods.  Stephan
and Horsch\cite{stephan}
studied the 2D $t$-$J$ model using the Lanczos algorithm
at $T=0$ on 16-20 site clusters.  They had relatively few {\bf k}-points
available due to the small cluster sizes, but their results are 
clearly consistent with a large Fermi surface.  The previous high
temperature series work\cite{singh}
presented results for a range of dopings at
$T=1.0J$, all of which are consistent with a large Fermi surface.
Quantum Monte Carlo has been used to study $n_{\bf k}$ in the Hubbard
model\cite{moreo}.  
Results for $n=0.87$, $U/t=4$, and $\beta t = 6$ on a $16\times16$
lattice are also consistent with a large Fermi surface.

Recently, the question of hole pockets in the Hubbard model has been
reexamined\cite{duffy}, 
with the suggestion that hole pockets would not show up
until fairly low temperatures of order $\beta t\sim 10$ for $U/t\gttrsim8$.
While the $t$-$J$ model and the Hubbard model are only directly
comparable for $J/t\ll 1$, our data at $J/t=0.5$ can be approximately
compared to the Hubbard data using the relation $J/t=4t/U$.  For
$J/t=0.5$ we expect results similar to those for the Hubbard model at
$U/t\approx 8$.  The temperature of our results at $J/t=0.5$ is
$T=0.2J=0.1t$.  This is low enough that we would expect to see an 
effect due to hole pockets of the type discussed for the Hubbard 
model\cite{duffy}.
Our current results do not show any indication of hole pockets of this
type.

In summary, we calculated a twelfth order high temperature series for
the momentum distribution $n_{\bf k}$ of the 2D $t$-$J$ model.  By 
extrapolating the series to $T=0.2J$ we examined $n_{\bf k}$ near
${\bf k}=(\pi/2,\pi/2)$, looking for hole pockets centered at this
wave vector.  We find a smooth monotonic wave vector dependence for
$n_{\bf k}$ through $(\pi/2,\pi/2)$, with no indication of hole pockets.
\bigskip

\noindent
{\it Acknowledgements.}
This work was partially supported by NSF grants DMR-9357199 (WOP) and
DMR-9616574 (RRPS). WOP thanks the ETH-Z\"urich for hospitality while
part of this work was being completed.\bigskip

\noindent
$^*$Current address.

\end{document}